\documentclass{Interspeech2024}
\usepackage{multirow}

\interspeechcameraready

\title{A Comprehensive Investigation on Speaker Augmentation for Speaker Recognition}

\name[affiliation={1,2}]{Zhenyu}{Zhou}
\name[affiliation={1}]{Shibiao}{Xu}
\name[affiliation={3}]{Shi}{Yin}
\name[affiliation={1}]{Lantian}{Li}
\name[affiliation={2}]{Dong}{Wang}

\address{
  $^1$School of Artificial Intelligence, Beijing University of Posts and Telecommunications, China \\
  $^2$Center for Speech and Language Technologies, BNRist, Tsinghua University, China \\
  $^3$Huawei Technologies Co., Ltd., China
  \thanks{This work was supported by the National Natural Science Foundation of China (NSFC) under Grants No.62301075/62171250.}}
\email{Corresponding authors:~lilt@bupt.edu.cn, Wangdong99@mails.tsinghua.edu.cn}


\begin{document}

\maketitle

\begin{abstract}

Data augmentation (DA) has played a pivotal role in the success of deep speaker recognition. Current DA techniques primarily focus on speaker-preserving augmentation, which does not change the speaker trait of the speech and does not create new speakers. Recent research has shed light on the potential of speaker augmentation, which generates new speakers to enrich the training dataset. In this study, we delve into two speaker augmentation approaches: speed perturbation (SP) and vocal tract length perturbation (VTLP). Despite the empirical utilization of both methods, a comprehensive investigation into their efficacy is lacking. Our study, conducted using two public datasets, VoxCeleb and CN-Celeb, revealed that both SP and VTLP are proficient at generating new speakers, leading to significant performance improvements in speaker recognition. Furthermore, they exhibit distinct properties in sensitivity to perturbation factors and data complexity, hinting at the potential benefits of their fusion. Our research underscores the substantial potential of speaker augmentation, highlighting the importance of in-depth exploration and analysis.

\end{abstract}

\section{Introduction}

In recent years, data augmentation (DA) techniques have exhibited profound efficacy in the realm of deep neural network-based speaker recognition~\cite{bai2021speaker,snyder2018x,zhu2019mixup,huang2019exploring,shahnawazuddin2020domain,wang2020investigation,dua2023data}, 
establishing themselves as indispensable for achieving good performance in numerous challenges and evaluations~\cite{sadjadi20222021,huh2023voxsrc,chen2022build,cnsrc2022tech}. 
The mainstream DA techniques involve manipulating speech samples by noise corruption~\cite{snyder2018x,tak2022rawboost} or spectrum masking~\cite{park2019specaugment,nam2022filteraugment}. These methods generate new training samples while preserving their speaker traits. 
The core idea of this type of DA methods is to increase the intra-speaker variations, 
enabling deep neural networks (DNNs) to extract more robust speaker-invariant patterns.

In contrast to speaker-preserving DA, recent studies have demonstrated the effectiveness of `\emph{speaker-expansion DA}' techniques, or speaker augmentation. Typical methods include speed perturbation (SP)~\cite{chen2022build,yamamoto19speaker,wang2020dku} and vocal tract length perturbation (VTLP)~\cite{wakamatsu2023vocal}. 
The purpose of this type of DA methods is to create new pseudo speakers, 
thereby encouraging the model to establish a better speaker embedding space and gain better discriminative capacity among speakers~\cite{snyder2016deep}. Compared to enriching utterances for each speaker, creating new speakers seems more crucial. This is because the number of speakers within a typical speech database is far from large. For instance, the popular VoxCeleb database contains less than 8,000 speakers. Considering the high dimensionality of typical speaker embeddings, for example, 256 in the x-vector paradigm~\cite{snyder2018x}, establishing a generalizable speaker embedding space with thousands of speakers is difficult. This is the reason why SP has yielded significant performance improvement in several benchmark evaluations~\cite{chen2022build,wang2020dku}.

Despite the remarkable contribution to the performance, it is surprising that there has not been a comprehensive study on speaker augmentation in the speaker recognition community. Their usage is largely motivated by the success of these methods in speech recognition~\cite{ko2015audio,jaitly2013vocal,cui2015data}, lacking a deeper understanding of how they work and in which condition they work. Moreover, SP is the only speaker augmentation method for a long time, with VTLP being picked up very recently~\cite{wakamatsu2023vocal}. Finally, while signal processing researchers have developed numerous tools for creating speech with diverse speaker characteristics, these tools have not yet been used for speaker augmentation. Overall, speaker augmentation appears to showcase a highly promising research direction that has garnered limited attention from researchers.

This paper takes one step further towards this promising direction. We investigate two speaker augmentation methods: time-domain SP~\cite{ko2015audio} and frequency-domain VTLP~\cite{jaitly2013vocal}. 
We conducted comprehensive studies on these two methods using two public datasets: VoxCeleb~\cite{nagrani2020voxceleb} and CN-Celeb~\cite{li2020cn}.
The experimental results show that both SP and VTLP are effective, though SP shows marginal superiority. Secondly, the two methods show different sensitivity to the perturbation factor and data complexity, leading to different behaviours when they are employed to perform speaker augmentation. Third, a combination of SP and VTLP may lead to further performance gains, with a carefully tuned fusion strategy.
To the best of our knowledge, this is the first effort to systematically investigate speaker augmentation methods within the realm of speaker recognition.


\section{Related Work}

Speaker augmentation, with the idea of creating pseudo speakers, perhaps originated from Jaitly and Hinton~\cite{jaitly2013vocal}. They proposed VTLP as an accompanying technique for the new generation of speech recognition systems based on deep neural nets. Intuitively, VTLP can be regarded as a `reverse operation' of Vocal Tract Length Normalization (VTLN)~\cite{lee1998frequency}, a popular approach to removing speaker variations. 
This normalization was crucial in the statistical model era, but in the deep learning era, neural nets are sufficiently powerful to deal with any speaker variation. Following this insight, VTLP was designed to create new speakers to enrich the training data, so the model can actively learn and handle variation related to speakers. 
Inspired by the success of VTLP, Tom etc.~\cite{ko2015audio} introduced SP as an alternative speaker augmentation method and demonstrated significant performance improvements with this method in speech recognition. Since then, SP and VTLP have been widely adopted in various speech processing tasks, including low-resource speech recognition~\cite{jaitly2013vocal,cui2015data,du2020speaker,geng2022investigation}, speech synthesis~\cite{cooper2020can}, 
speech separation~\cite{wang2023speakeraugment}, and speaker recognition~\cite{yamamoto2019speaker,chen2022build,wang2020dku,wakamatsu2023vocal}.

In the realm of speaker recognition, SP has been widely acknowledged as a pivotal technique in numerous evaluations~\cite{chen2022build,wang2020dku}. Nevertheless, the exploration of VTLP remains limited. Recently, Tomoka et al.~\cite{wakamatsu2023vocal} experimented on VTLP-based speaker augmentation. Their findings revealed that solely applying VTLP did not lead to performance enhancements. However, when coupled with a selection process to choose `good' generations and combined with noise augmentation, some improvements were observed. 
However, their study was conducted using a non-standard dataset, and a comparison with SP was absent.


\section{Review for SP and VTLP}

\subsection{Speed Perturbation (SP)}

Speed perturbation operates by resampling the speech signal within the time domain. For a given speech signal $x(t)$, a perturbation factor $\alpha$ is applied to resample the signal along the time axis, yielding the output $y(t)$: 

\begin{equation}
\label{eq:rcr}
y(t) = x(\alpha t).
\end{equation}

\noindent This operation in the time domain corresponds to a specific transformation in the frequency domain, described by the following change:

\begin{equation}
\label{eq:rcr}
X(f) \rightarrow \frac{1}{\alpha}X(\frac{1}{\alpha}f),
\end{equation}

\noindent where $X(f)$ and $\frac{1}{\alpha}X(\frac{1}{\alpha}f)$ represent the Fourier transform of $x(t)$ and $y(t)$, respectively. 
A key observation is that SP not only changes the duration of the speech signal but also modifies the spectrum by stretching or compressing the spectral range. Two important changes caused by the spectral modification are: (1) the fundamental frequency (F0) is increased (downsampling) or decreased (upsampling); (2) the spectrum envelope is stretched (downsampling) or compressed (upsampling), leading to the change of the location and width of the formants (F1, F2, ...). All these alterations lead to deviations in speaker characteristics, akin to creating a new speaker.

\subsection{Vocal Tract Length Perturbation (VTLP)}


The length of the vocal tract varies across different speakers, leading to unique traits for each individual~\cite{wakita1973direct}. 
The core idea of VTLP is to modify an existing speech signal to simulate the change in its vocal tract length.

For a speech signal $x(t)$, its spectrum $X(f)$ can be obtained through the Fourier transform. VTLP employs a piece-wise linear warping in the frequency domain to approximate the change in the vocal tract length, as defined below~\cite{jaitly2013vocal}:

\begin{equation}
f' = \begin{cases}
      \begin{aligned}
        & \alpha f               && 0 \leq f \leq f_0, \\
        & \frac{f_\text{max}-\alpha f_0}{f_\text{max}-f_0}(f-f_0)+\alpha f_0  && f_0 \textless f \leq f_\text{max}.
      \end{aligned}
     \end{cases}
\end{equation}

\noindent where $\alpha$ is the perturbation factor, $f_\text{max}$ is the maximum sampling frequency and $f_0$ is a boundary frequency chosen such that the warping function $\alpha f$ covers significant formats. In this study, $f_\text{max}$ is 8,000 and $f_0$ is set to 4,800. 

\subsection{SP vs. VTLP}

Theoretically, both SP and VTLP can be viewed as warping functions in the frequency domain, albeit with distinct forms of warping functions. Another differentiating factor is that SP alters the speaking rate, introducing additional variations in speaking styles by shortening or elongating the duration of speech patterns. Additionally, a noteworthy characteristic of SP is that when the speed is decreased (i.e., $\alpha$ $\textless$ 1), the energy of the speech signal concentrates in low-frequency regions near zero~\cite{ko2015audio}, a feature not observed in VTLP.

Given these distinctions, it is crucial to conduct a careful comparative study to analyze the behaviour of these methods in creating new speakers. Furthermore, exploring whether these two methods can be combined to achieve additional performance enhancements represents an intriguing avenue that has yet to be explored.

\section{Experiments}

In this section, we present comprehensive experiments to evaluate and compare the effectiveness of SP and VTLP. The objective of these experiments is to examine the specialities of these two methods and provide an in-depth understanding of their practical utility.

\subsection{Data}

The datasets used in our experiments are VoxCeleb1~\cite{nagrani2017voxceleb} and CN-Celeb1~\cite{fan2020cn}. 
VoxCeleb1 encompasses a total of 1,251 speakers, including a development set VoxCeleb1.dev for model training and an evaluation set VoxCeleb1.eval for performance assessment. CN-Celeb1 comprises 998 speakers in total, including a development set CN-Celeb1.dev for model training and an evaluation set CN-Celeb1.eval for testing.

It is noteworthy that in our experiments, we did not select the larger VoxCeleb2 and CN-Celeb2 datasets because we wanted to simulate a training condition with a limited number of speakers, thus more clearly exposing the benefits of the speaker augmentation methods. Furthermore, the MUSAN dataset~\cite{snyder2015musan} was used to perform speaker-preserving DA. This is intended to construct more varied acoustic conditions.

\subsection{Settings}

We follow the cnceleb/v2 recipe of the Sunine toolkit\footnote{https://gitlab.com/csltstu/sunine/} to construct the deep speaker model. The structure of the model is ResNet34 with squeeze-and-excitation (SE) layers, accompanied by an attentive statistics pooling layer to produce utterance-level representations. These representations are then transformed by a fully connected layer to obtain a 256-dimensional x-vector. The model is trained by the Adam optimizer with AAM-softmax~\cite{deng2019arcface} as the training objective. The simple cosine distance is used to score the trials.

\subsection{Deviation Analysis}

\subsubsection{Principle of speaker augmentation}

In various tasks, such as speech recognition, speaker augmentation typically carries no inherent risks. However, this is not the case in the context of speaker recognition. This is because the generated speech will be assigned a new speaker label, and the accuracy of this labelling is paramount to the outcome. Clearly, the accuracy of the labels is related to how the generated speech deviates from the original speech in terms of speaker traits: the larger the deviation, the more accurate the labels.

There are three key criteria for assessing effective speaker augmentation: (1) the generated speech should not be excessively distorted and remain intelligible human speech; (2) the generated speech should exhibit clear differentiation in speaker characteristics from the original speech; (3) speech generated from the same speaker should demonstrate consistency and similarity. These criteria are all linked to the intensity of perturbation, which is controlled by the perturbation factor $\alpha$ for both SP and VTLP. Hence, our focus is on investigating the influence of $\alpha$ on the generated speech using these two methods.

\subsubsection{No-distortion range}

We initiated a human evaluation by listening to the generated speech. Our findings indicated that for both SP and VTLP, the optimal range for the perturbation factor $\alpha$ is between 0.8 and 1.2 to avoid noticeable distortion. Based on this observation, we have set $\alpha$ to the range of 0.8-1.2 for all subsequent analyses.

\subsubsection{Deviation distribution curve}

In this step, we assessed the deviation in speaker traits when a speech utterance was perturbed by SP or VTLP. The deviation in speaker traits was quantitatively assessed by calculating the cosine distance between the speaker embeddings of the speech signal before and after the perturbation. The specific steps involved in this evaluation are outlined below.

\begin{itemize}

\item For each utterance $u_{si}$ in VoxCeleb1.dev and CN-Celeb1.dev, where $s$ and $i$ index the speaker and utterance respectively, apply SP or VTLP to modify the speech, resulting in a new utterance $u'_{si}$.

\item Utilize a pre-trained speaker embedding model to derive the speaker embeddings $e_{si}$ and $e'_{si}$ for the utterances $u_{si}$ and $u'_{si}$.

\item Compute the cosine distance between each pair of embeddings $(e_{si}, e'_{si})$ to determine the utterance-level deviation. Aggregate the utterance-level deviations to compute the speaker-level deviation, denoted as $d_{s}=\sum_i 1$-$\text{cos}(e_{si}, e'_{si})$.

\item Plot a deviation distribution curve where the x-axis represents the speaker-level deviation and the y-axis illustrates the proportion of speakers experiencing a specific deviation.

\end{itemize}

\begin{figure}[htbp]
  \centering
  \includegraphics[width=0.5\textwidth]{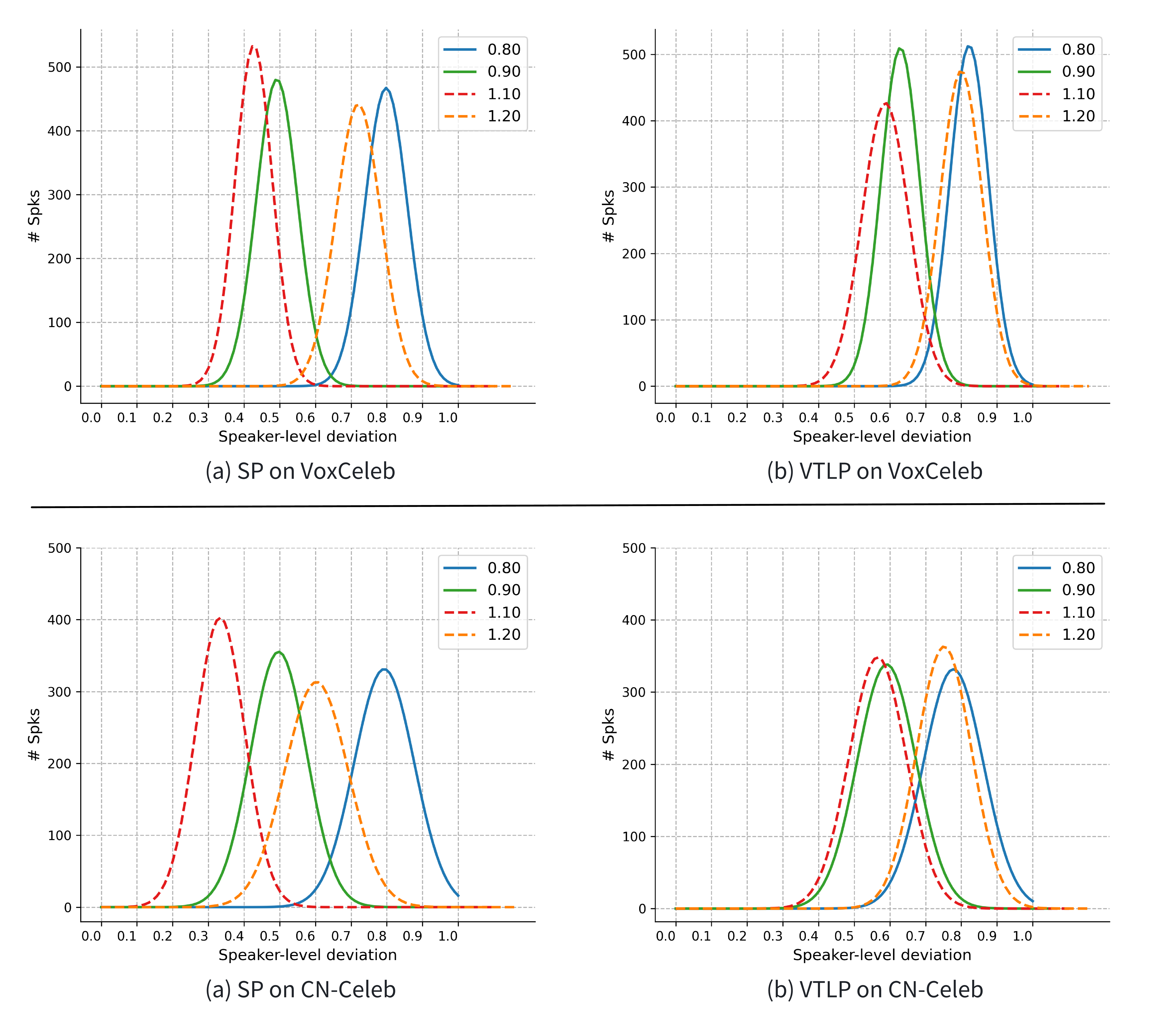}
  \caption{The deviation distribution curves. Curves with $\alpha>1$ are plotted in dotted lines, while curves with $\alpha<1$ are plotted in solid lines.}
  \label{fig:speaker}
\end{figure}

Figure~\ref{fig:speaker} presents deviation distribution curves with different settings of $\alpha$. Each plot corresponds to a specific perturbation approach (SP/VTLP) applied to a particular dataset (VoxCelb/CN-Celeb). Key observations are as follows.

\begin{itemize}

\item For both SP and VTLP, a $\alpha$ value further from 1 results in a greater deviation compared to a $\alpha$ value closer to 1, as evidenced by a rightward shift on the distribution curve. This outcome is to be expected.

\item For both SP and VTLP, deviations induced by a specific $\alpha$ form a compact distribution with a variance below 0.01. Furthermore, the distributions corresponding to different $\alpha$ values are distinctly separated. These findings suggest the feasibility of generating additional pseudo speakers by aggregating perturbed speech with varying $\alpha$ values.

\item In comparison to VTLP, the impact of SP on deviation is more responsive to the $\alpha$ value, evident in the minimal overlap between the pair of deviation distribution curves corresponding to (1.10, 1.20) or (0.80, 0.90).


\item A comparison between VoxCeleb and CN-Celeb reveals a greater distribution curve overlap for both SP and VTLP in CN-Celeb. This could be attributed to the more complex acoustic conditions and speaking styles present in CN-Celeb. The behaviour of any perturbation method may vary inconsistently with such complex data, resulting in dispersed deviation distributions. This observation underscores the necessity for meticulously designed perturbations for complex data, and speech produced by different perturbation factors should not be simply regarded as two different persons. It also suggests the need for the development of selection strategies, as discussed in~\cite{wakamatsu2023vocal}.

\end{itemize}

\subsubsection{Deviation-Perturbation curve}

\begin{figure}[htbp]
  \centering
  \includegraphics[width=0.5\textwidth]{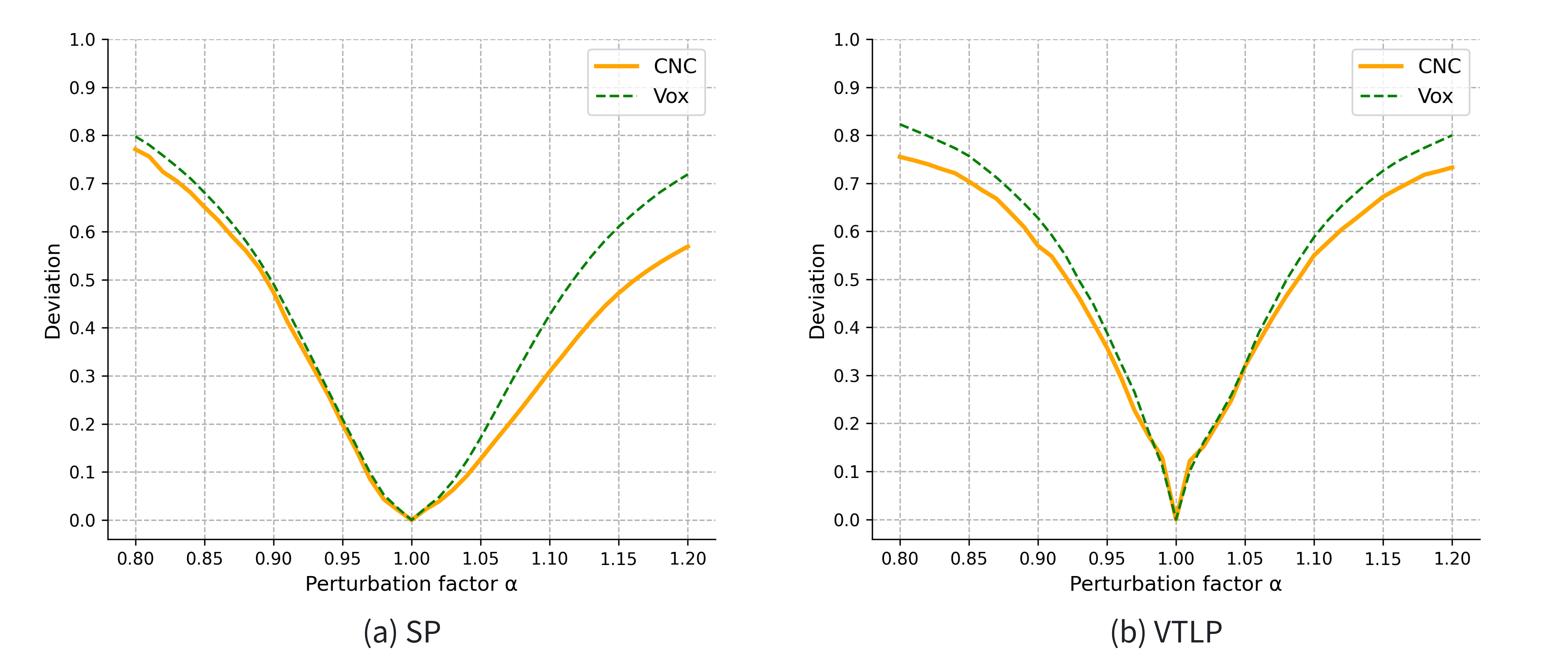}
  \caption{The deviation-perturbation curve with (a) SP and (b) VTLP. }
  \label{fig:global}
\end{figure}

We plot a deviation-perturbation curve that reflects 
the average deviation across all speakers against $\alpha$ values, ranging incrementally from 0.8 to 1.2, as shown in Figure~\ref{fig:global}. These curves provide a clearer comparison of the influence of $\alpha$ on SP and VTLP.

It is evident that as $\alpha$ deviates from 1 in both directions, the curves for VTLP exhibit a noticeable symmetry on both VoxCeleb and CN-Celeb datasets. This symmetry is less pronounced for SP, particularly on the more complex CN-Celeb dataset, aligning with the findings in Fig.~\ref{fig:speaker}. Additionally, the distribution curves for SP appear smoother than those for VTLP, suggesting that VTLP may induce more rapid changes in speaker traits. This factor should be considered when utilizing these methods to generate new speakers, especially when creating multiple speakers from the same individual using different $\alpha$ values.


\subsection{Speaker Recognition Results}

\subsubsection{Results with SP and VTLP}

This section presents the speaker recognition results with speaker augmentation by SP and VTLP. Table~\ref{tab:sp} and Table~\ref{tab:vtlp} present the performance with SP and VTLP respectively, and in each table, results on both VoxCeleb1 and CN-Celeb1 are reported, with various configurations of $\alpha$.

\begin{table}[h]
\centering
\caption{Results in EER(\%) with SP-based speaker augmentation.}
\label{tab:sp}
\resizebox{1.0\columnwidth}{!}{
\begin{tabular}{l|c|c|cc}
\toprule
\multirow{2}{*}{Method} & \multirow{2}{*}{$\alpha$} & \multirow{2}{*}{\# Spks} & \multicolumn{2}{c}{EER(\%)} \\
                        &            &            &  Vox   &   CNC      \\
\midrule
Baseline                & -          & $\times$1  & 2.457  &   11.067   \\
\midrule
\multirow{5}{*}{SP}     & 0.95, 1.05 & $\times$3  & 2.776 & 12.171      \\
                        & 0.90, 1.10 & $\times$3  & \textbf{1.760} & 10.307 \\
                        & 0.85, 1.18 & $\times$3  & 1.776 & 9.592 \\
                        & 0.80, 1.20 & $\times$3  & 1.824 & \textbf{9.327} \\
\cmidrule(r){2-5}
                        & 0.80, 0.90, 1.10, 1.20  & $\times$5 & \textbf{1.627} & 9.868 \\
\bottomrule
\end{tabular}}
\end{table}

\begin{table}[h]
\centering
\caption{Results in EER(\%) with VTLP-based speaker augmentation.}
\label{tab:vtlp}
\resizebox{1.0\columnwidth}{!}{
\begin{tabular}{l|c|c|cc}
\toprule
\multirow{2}{*}{Method} & \multirow{2}{*}{$\alpha$} & \multirow{2}{*}{\# Spks} & \multicolumn{2}{c}{EER(\%)} \\
                        &            &            &  Vox   &   CNC      \\
\midrule
Baseline                & -          & $\times$1  & 2.457  &   11.067   \\
\midrule
\multirow{6}{*}{VTLP}   & 0.93, 1.07 & $\times$3  & 2.080  & 11.276  \\
                        & 0.90, 1.10 & $\times$3  & \textbf{2.010}  & 10.397  \\
                        & 0.83, 1.17 & $\times$3  & 2.069    & \textbf{10.115}   \\
                        & 0.80, 1.20 & $\times$3  & 2.149  & 10.234  \\
\cmidrule(r){2-5}                        
                        & 0.80, 0.90, 1.10, 1.20  & $\times$5 & \textbf{1.883} & \textbf{9.930} \\
\bottomrule
\end{tabular}}
\end{table}

We observe that: (1) With an appropriate setting of $\alpha$, 
both SP and VTLP significantly enhance performance compared to the baseline system. This underscores the effectiveness of the two speaker augmentation methods. (2) Comparing SP and VTLP, SP generally outperforms VTLP. This may be attributed to the complex effect of SP on both speed and spectral warping. 
(3) The choice of perturbation factor $\alpha$ is critical. 
Setting $\alpha$ close to 1 is risky and may result in a severe performance drop. This is not surprising as this setting may lead to 
insufficient speaker deviation so the perturbed data should not be regarded as a new speaker. In contrast, setting $\alpha$ to far from 1 is also not optimal, probably due to the potential distortion caused by the strong perturbation. (4) Comparing VoxCeleb and CN-Celeb, VoxCeleb requires a smaller $\alpha$ to reach optimal performance, while CN-Celeb demands a larger $\alpha$. This observation is consistent with the phenomena illustrated in Figure~\ref{fig:speaker} and Figure~\ref{fig:global}: both figures show that the deviation on the speech of VoxCeleb is more sensitive to the change of $\alpha$.

The key observation is the ability to generate more than two new speakers by varying $\alpha$ values. For example, when producing four new speakers using VTLP with four distinct $\alpha$ values (0.80, 0.90, 1.10, 1.20), improved performance was achieved on both VoxCeleb and CN-Celeb compared to generating only two new speakers. While a similar performance enhancement was observed for SP on VoxCeleb, the same effect was not observed on CN-Celeb. These findings align perfectly with the deviation-perturbation curves depicted in Figure~\ref{fig:global}, where the VTLP curves exhibit sharper delineations, allowing for segmentation into multiple segments with varying deviations. This segmentation facilitates the creation of diverse new speakers. Conversely, the SP curves are smoother and less conducive to segmentation. Particularly on CN-Celeb, the downward slope on the right side of the LP curve indicates reduced differentiation among speakers with varying $\alpha$ values.

\subsubsection{Results of SP + VTLP}

We combine SP and VTLP to get more pseudo speakers, by choosing the best \#Spks$\times$3 configuration for SP and VTLP shown in Table~\ref{tab:sp} and~\ref{tab:vtlp} respectively, and pool the created data to train the model. The results are reported in Table~\ref{tab:fusion}.

\begin{table}[h]
\centering
\caption{Results in EER (\%) with SP + VTLP.}
\label{tab:fusion}
\resizebox{0.85\columnwidth}{!}{
\begin{tabular}{l|c|c|cc}
\toprule
\multirow{2}{*}{Method} & \multirow{2}{*}{$\alpha$} & \multirow{2}{*}{\# Spks} & \multicolumn{2}{c}{EER(\%)} \\
                        &            &            &  Vox   &   CNC      \\
\midrule
Baseline                & -          & $\times$1  & 2.457  & 11.067 \\
\midrule
SP                      & 0.90, 1.10 & $\times$3  & 1.760  & -      \\
VTLP                    & 0.90, 1.10 & $\times$3  & 2.010  & -      \\
SP + VTLP               & Fusion     & $\times$5  & \textbf{1.744}    & -      \\
\midrule                
SP                      & 0.80, 1.20 & $\times$3  & -      & \textbf{9.327}  \\
VTLP                    & 0.83, 1.17 & $\times$3  & -      & 10.115  \\
SP + VTLP               & Fusion     & $\times$5  & -      & 9.665    \\
\bottomrule
\end{tabular}}
\end{table}

The results reveal that combining the two perturbation methods leads to further performance improvements on the VoxCeleb dataset, but not as good as the SP perturbation on the CN-Celeb dataset. This suggests that the two augmentation methods exhibit complementarity, but a more careful exploration of the fusion scheme is required.

\section{Conclusion}

This study delves into a comprehensive analysis of two prominent speaker augmentation techniques, Speed Perturbation (SP) and Vocal Tract Length Perturbation (VTLP), within the realm of speaker recognition tasks. Through extensive analyses conducted on the VoxCeleb1 and CN-Celeb1 datasets, we unveil distinct properties of SP and VTLP. In addition to explaining the speaker recognition outcomes using these augmentation methods, our analysis offers insights into the development of safer and more diverse speaker augmentation techniques, in particular the creation of more distinct speakers from a single speaker. We also observed the feasibility of combining SP and VTLP, highlighting the importance of carefully designing the fusion approach. Future research endeavours will focus on exploring optimal fusion strategies for SP and VTLP, as well as analyzing and employing other speaker augmentation methods such as various voice morphing and conversion techniques.


\newpage

\bibliographystyle{IEEEtran}
\bibliography{mybib}

\end{document}